\documentclass[
 reprint,
nofootinbib,
 amsmath,amssymb,
 aps,
 twocolumn
]{revtex4}

\usepackage{graphicx}
\graphicspath{{/paper/}}
\usepackage{dcolumn}% Align table columns on decimal point
\usepackage{bm}% bold math
\usepackage{slashed}
\usepackage{enumitem}
\usepackage{blindtext} 
\usepackage[linktocpage]{hyperref}
\usepackage[svgnames]{xcolor}
\hypersetup{
     colorlinks = true,
     linkcolor = blue,
     anchorcolor = blue,
     citecolor = magenta,
     filecolor = blue,
     urlcolor = blue
     }

\begin{document}
\preprint{APS/123-QED}

\title{A New Look at the $C^{0}$-formulation of the Strong Cosmic Censorship Conjecture}

\author{Aditya Iyer}
\email{aditya.iyer@physics.ox.ac.uk}
\affiliation{Department of Physics, University of Oxford, Parks Road, Oxford, OX1 3PU, UK}

\author{Alexander Y. Yosifov}
\email{alexanderyyosifov@gmail.com}
\affiliation{Space Research and Technology Institute, Bulgarian Academy of Sciences, Akad. G. Bonchev Street, Building 1, Sofia, 1113, Bulgaria}

\author{Vlatko Vedral}
\email{vlatko.vedral@physics.ox.ac.uk}
\affiliation{Department of Physics, University of Oxford, Parks Road, Oxford, OX1 3PU, UK}
\affiliation{Centre for Quantum Technologies, National University of Singapore, 3 Science Drive 2, 117543 Singapore, Singapore}
\affiliation{
 Department of Physics, National University of Singapore, 2 Science Drive 3, Singapore 117542}
\date{\today}

\date{\today}% It is always \today, today,

\begin{abstract}
We examine the $C^{0}$-formulation of the strong cosmic censorship conjecture (SCC) from a quantum complexity-theoretic perspective and argue that for generic black hole parameters as initial conditions for the Einstein equations, corresponding to the expected geometry of a hyperbolic black hole, the metric is $C^{0}$-extendable to a larger Lorentzian manifold across the Cauchy horizon. To demonstrate the pathologies associated with a hypothetical validity of the $C^{0}$ SCC, we prove it violates the "complexity=volume" conjecture for a low-temperature hyperbolic AdS$_{d+1}$ black hole dual to a CFT living on a ($d-1$)-dimensional hyperboloid $H_{d-1}$, where in order to preserve the gauge/gravity duality we impose a lower bound on the interior metric extendability of order the classical recurrence time. 

\begin{description}
\item [{PACS~numbers:}] 04.70.Dy, 03.67.a, 04.20.Dw{\small \par}
\item [{Keywords:}] quantum complexity, black holes, strong cosmic censorship{\small \par}
\end{description}
\end{abstract}


\maketitle

%\tableofcontents

\section{Introduction}

Black holes have large interiors \cite{rovelli,yeng}.
The interior volume of a black hole of mass $M$ is $\sim M^{5}$, while its lifespan is $\sim M^{3}$. As was shown in \cite{rovelli} for the case of a one-parameter static spherically-symmetric solution to the Einstein vacuum equations $R_{\mu\nu}=0$ (\textit{i.e.} Schwarzschild black hole), even though the horizon area is time translation invariant for an asymptotic observer, its interior volume, defined in a geometrically covariant way as the maximal spacelike hypersurface bounded by a two-sphere foliating the horizon, is dynamical, and keeps growing. Where according to classical general relativity, the interior volume continues to grow unbounded. For the simplest case of a Schwarzschild metric, expressed in advanced Eddington-Finkelstein coordinates 

\begin{equation}
ds^{2}=-\left(1-\frac{2M}{r}\right)dv^{2}+2dvdr+r^{2}\sin^{2}\theta d\phi^{2}+d\theta^{2}
\end{equation}
where $v=t+\int\frac{dr}{1-2M/r}$, and $t$ is Schwarzschild time,
the interior volume grows linearly in the advanced time $v$, where
up to a prefactor of $3\sqrt{3}\pi$ it reads 

\begin{equation}
\label{eq:v(v)}
V(v)=M^{2}v
\end{equation}
The maximum of Eq. (\ref{eq:v(v)}) is obtained by extremizing the integral

\begin{equation}
\label{eq:integralV}
V=\int^{v}\int_{S^{2}}\left[r^{2}\left(\frac{2M}{r}-1\right)^{3/2}\right]\sin\theta d\theta d\phi dv
\end{equation}
which maximizes at $r_{\text{max}}=3M/2$ and $S^{2}$ denotes the two-sphere foliating the horizon.\\
This non-trivial relation between horizon area and interior volume was further examined in Ref. \cite{lorenzo} where it was argued that for a Schwarzschild black hole the volume of the maximal bounded spacelike hypersurface behind the horizon continues to grow until the black hole reaches the Planck scale. Without loss of generality, the results were expanded \cite{jakobsson} for the rotating $a\neq0$ subextremal $|a|<M$ two-parameter family of stationary axisymmetric solutions in asymptotically flat spacetime (\textit{i.e.} Kerr black hole). Where assuming $M\gg1$, and the metric curvature is weak at both $r=2M$ and $r=3M/2$, then similar reasoning to that in $a=0$ Schwarzschild  geometry was shown to be valid. Namely, for a subextremal Kerr black hole the leading contribution to the interior volume is due to the maximal Cauchy development of initial data defined on a spacelike hypersurface which lies just barely within the black hole.\\
This growth of the interior region was later quantified in the context of AdS/CFT, where the volume of a black hole in $d$-dimensional asymptotically AdS spacetime was conjectured to be dual to the complexity of a thermal state CFT living on the $(d-1)$-dimensional AdS boundary, hence "complexity=volume" (CV).\footnote{Interestingly, it was then argued in Ref. \cite{susskind} that complexity, similar to entropy, admits a Second law.}\\
Lately, extendability of the interior metric of dynamical black holes with inner horizon has attracted much attention as it has been demonstrated \cite{dafermos,dafermos2,dafermos3,chesler} that for a maximal Cauchy development of suitable initial data, considered on a spacelike hypersurface $\Sigma$ inside a subextremal Kerr (or Reissner-Nordstrom) black hole, the metric is $C^{0}$-stable, meaning it is \textit{extendable} to a larger Lorentzian manifold $\tilde{\mathcal{M}}\setminus\mathcal{M}$ across a non-trivial part of the Cauchy horizon. Hence, invalidating the $C^{0}$ form of Penrose's celebrated SCC.\\
In the main part of this paper, building on Ref. \cite{dafermos}, we revisit the $C^{0}$-formulation of the SCC from a quantum complexity-theoretic perspective. We investigate the $C^{0}$ SCC in the low-temperature limit of a hyperbolic AdS$_{d+1}$ black hole dual to a CFT living on a $(d-1)$-dimensional
hyperboloid to demonstrate that if the metric terminates at the Cauchy horizon, then the CV-duality (\textit{i.e.} gauge/gravity correspondence) will be violated at late times. In light of this, we propose a novel conjecture relating the $C^{0}$-stability of the metric to the complexification of the system, and suggest the extendability of the black hole interior is bounded from below by the classical recurrence time.\\
The paper is organized as follows. In Section \ref{sec:section2} we review the Kerr geometry, and focus on the role of the Cauchy horizon. We then look at the $C^{0}$-formulation of Penrose's SCC and its implications for the interior Kerr black hole region. In Section \ref{sec:section3} we examine the implications of the $C^{0}$-instability of the metric for Kerr-AdS spacetime. Section \ref{sec:Section4} is dedicated to the study of the inconsistencies in the bulk/boundary duality yielded by a hypothetical inextendability of the interior metric beyond the Cauchy horizon. 

\section{The Kerr Metric and $C^{0}$-inextendability}
\label{sec:section2}

The subextremal $\left|a\right|<M$ Kerr family of solutions describes a metric whose line element, in Eddington-Finkelstein coordinates, reads

\begin{multline}
ds^{2}=-\frac{\Delta-a^{2}\sin^{2}\theta}{\rho^{2}}dv^{2}+2dvdr+\rho^{2}d\theta^{2}+\frac{A\sin^{2}\theta}{\rho^{2}}d\phi^{2}\\
-2a\sin^{2}\theta drd\phi-\frac{4aMr}{\rho^{2}}\sin^{2}\theta dvd\phi
\end{multline}
where

\begin{equation}
\begin{aligned}\Delta\equiv r^{2}-2Mr+a^{2}\\
\rho^{2}\equiv r^{2}+a^{2}\cos^{2}\theta\\
A\equiv(r^{2}+a^{2})^{2}-a^{2}\Delta\sin^{2}\theta
\end{aligned}
\end{equation}
Here, $a=J/M$ is the rotation parameter which gives the spin-to-mass
ration of the black hole, where the extremal case corresponds to $a\sim\mathcal{O}(1)$. The metric admits two horizons, respectively at $r_{\pm}=M\pm\sqrt{M^{2}-a^{2}}$,
where $r_{-}$:=$\{{{{r}}}=\mathcal{C}\mathcal{H}^{+}\}$ is the inner (Cauchy) horizon and $r_{+}$:=$\{{{{r}}}=\mathcal{H}^{+}\}$ is the outer (event) horizon.\footnote{Similar notation will be employed in Section \ref{sec:section3} as well.} Namely the Cauchy horizon will be the focus of the rest of the paper.\\
In Kerr geometry the maximal Cauchy development of generic asymptotically flat initial data, defined on a spacelike hypersurface $\Sigma$ within the black hole (\textit{i.e.} the non-empty complement of $\mathcal{M}\setminus J^{-}(\mathcal{I}^{+})$), is \emph{non-uniquely }extendable as a solution to the vacuum Einstein equations. This means lack of future determinism for all infalling timelike geodesics $\gamma$ which reach $\tilde{\mathcal{M}}\setminus\mathcal{M}$. Thus, the region beyond the Cauchy horizon $\mathcal{C}\mathcal{H}^{+}$ cannot be uniquely determined, implying a breakdown of global hyperbolicity of the metric, as predicted by classical general relativity. Upon maximal evolution of the initial data on the spacelike hypersurface $\Sigma$ to a larger Lorentzian manifold $\tilde{\mathcal{M}}\setminus\mathcal{M}$, $\Sigma$ will no longer be the Cauchy hypersurface for $\tilde{\mathcal{M}}$ which extends across $\mathcal{C}\mathcal{H}^{+}$.\\
This globally non-hyperbolic future development of the initial data found in Kerr (and also Reissner-Nordstrom) solutions is in sharp contrast to the simple one-parameter Schwarzschild subfamily of solutions, where the maximal Cauchy evolution is future \emph{inextendable} as a $C^{0}$ Lorentzian manifold across $\mathcal{C}\mathcal{H}^{+}$. In Schwarzschild the future is uniquely determined, namely for $r<r_{+}$ all timelike geodesics $\gamma$ are incomplete, and asymptote to a spacelike boundary in finite proper time.\\
In an attempt to preserve the global hyperbolicity of the metric Penrose proposed \cite{penrose} the so-called \emph{strong cosmic censorship conjecture} which, generally, reads\\

\textbf{Conjecture \hypertarget{conjecture}{1} (the $C^{0}$-form of the SCC):} The maximal Cauchy evolution of generic asymptotically flat (AdS) vacuum initial data is \textit{inextendable} to a larger $C^{0}$ Lorentzian manifold $\tilde{\mathcal{M}}$ across the Cauchy horizon $\mathcal{C}\mathcal{H}^{+}$.\\

The role of the $C^{0}$-formulation of the SCC is to bring uniqueness to the solutions of the vacuum Einstein equations, and hence make the future of all infalling timelike geodesics deterministic. The $C^{0}$-inextendability conjecture restores the predictability of the vacuum equations inside dynamical black holes by introducing a spacelike finite boundary across which the metric cannot be extended even as a (continuous) $C^{0}$.\\
Recent advancements, however, see Refs. \cite{dafermos,luk2}, suggest Conjecture \hyperlink{conjecture}{1} is \textit{false}, where the main results relevant to us can be summarized by the following theorem\\

\textbf{Theorem \hypertarget{theorem}{1} (Dafermos-Luk):} Consider general vacuum initial data corresponding to the expected induced geometry of a dynamical black hole settling down to Kerr (with parameters $0<\left|a\right|<M$) on a suitable spacelike hypersurface $\Sigma_{0}$ in the black hole interior. Then the maximal future development of the spacetime $(\mathcal{M},g)$ corresponding to $\Sigma_{0}$ is globally covered by a double null foliation and has a non-trivial Cauchy horizon $\mathcal{C}\mathcal{H}^{+}$ across which the metric is continuously extendable.\\

The theorem suggests that general exterior subextremal spacetimes, emerging from initial data \textit{sufficiently} close to Kerr data, have $C^{0}$-stable non-trivial Cauchy horizons in their interiors. This, together with the global existence statement in the exterior (weak cosmic censorship), are already a strong argument \textit{against} Conjecture \hyperlink{conjecture}{1}. In fact, the weak cosmic censorship (\textit{i.e.} the completeness of null infinity $\mathcal{I}^{+}$) makes an even stronger statement as it implies the entire Penrose diagram is nonlinearly stable for solutions to $R_{\mu\nu}=0$. So given \textit{stability} in the Kerr exterior $J^{-}(\mathcal{I}^{+})$, we can infer similar properties to hold in the non-empty complement of $\mathcal{M}\setminus J^{-}(\mathcal{I}^{+})$ as well.\\ 
Moreover, $C^{0}$-stability results of $\mathcal{C}\mathcal{H}^{+}$ have also been derived for a broader class of systems, \textit{e.g.} for Reissner-Nordstrom solutions to the Einstein-Maxwell-real-scalar-field equations \cite{dafermos2,dafermos3,chesler}, while in Ref. \cite{moortel}, assuming \textit{polynomial} decay rate, for suitable initial data of the Einstein-Maxwell-Klein-Gordon equations in a Reissner-Nordstrom background, the spacetime was shown to be continuously extendable.\footnote{It should be noted, however, that although the metric may admit continuous $C^{0}$-extendability across the Cauchy horizon, it may still be \emph{inextendible} as a $C^{2}$ Lorentzian manifold, assuming higher regularity conditions are imposed. Hence, the Christoffel symbols can still fail to be square integrable.} On a linear level, the $C^{0}$ SCC, being the strongest inextendability statement, has been shown not to admit a blow up at $\mathcal{C}\mathcal{H}^{+}$ even for the simpler Kerr or Reissner-Nordstrom solutions in asymptotically flat spacetime, given that the polynomially decaying scalar waves solving the Klein-Gordon equation propagate to the black hole interior. Similarly, boundedness at $\mathcal{C}\mathcal{H}^{+}$ has also been demonstrated \cite{ds} for Kerr (and Reissner-Nordstrom)-dS spacetime with the decay rate here being exponential.\\
Despite the growing interest in the subject of spacetime extendability in black hole interiors, all $C^{0}$ SCC formulations rely on classical geometric arguments and scalar wave decay rate estimations, making the SCC one of the few big problems in gravity yet to receive the quantum treatment. As such, Sections \ref{sec:section3} and \ref{sec:Section4} constitute the first application of quantum complexity in this context.

\section{Quantum Complexity and the $C^{0}$-stability}
\label{sec:section3}

We begin this Section with a brief review of the CV-duality and outline a newly-developed mechanism for measuring it in terms of the flux of a conserved volume current. Then, we discuss what the $C^{0}$-formulation of the SCC (Conjecture \hyperlink{conjecture}{1}) implies for the interior of a Kerr-AdS black hole. Although some of the original results were derived for the case of an eternal two-sided Schwarzschild-AdS black hole, Fig. \ref{fig:fig3a}, they have been shown to be valid for a large class of solutions.

\subsection{Quantum Complexity}
\begin{figure}
\includegraphics[scale=0.43]{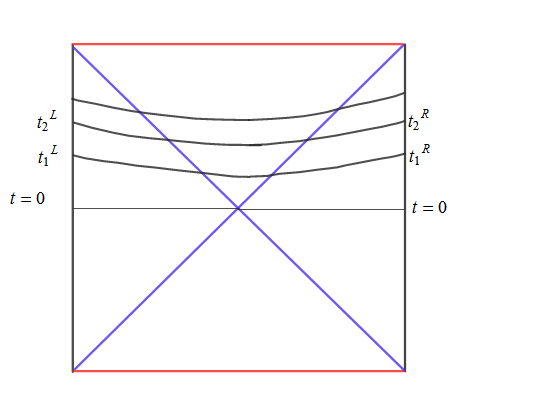}\caption{A diagram of an eternal two-sided Schwarzschild-AdS black hole. The black lines, denoted by $t$, describe the boundary Cauchy foliation by slices which are orthogonal to an asymptotic Killing flow yielding time translations, the blue lines indicate the horizons, and the red lines are the singularities. Throughout the paper we will only focus on the upper ($t>0$) part of the diagram.}
\label{fig:fig3a}
\end{figure}

Quantum complexity increases linearly for time exponential in the entropy $t_{\text{max}}\sim e^{S}$, known as the classical recurrence time. At this point it saturates and remains at its maximum $\mathcal{C}_{\text{max}}=e^{S}$ for $t_{\text{qr}}\sim e^{e^{S}}$, known as the quantum recurrence time, and then fluctuates back to its initial low value \cite{susskind2}.\\
The CV-duality \cite{susskind3} suggests the complexity of the boundary CFT is dual to the volume of the maximal spacelike hypersurface (\emph{i.e.} codimension-one submanifold)  which asymptotes to some time, anchored at the conformal AdS boundary, see Fig. \ref{fig:fig3a}

\begin{equation}
\label{eq:C=V}
\mathcal{C}\sim\frac{V}{\hbar G_{N}l}
\end{equation}
where $l$ is the characteristic AdS length scale, $\hbar=h/2\pi$ is the reduced Planck constant, and $G_{N}$ denotes Newton's constant, where we assume $c=\hbar=G_{N}=1$.

The CV-duality has been shown to hold for a wide class of theories and in arbitrary number of dimensions. Its validity has been demonstrated for the simplified setting of a (1+1)-dimensional Reissner-Nordstrom background in JT gravity \cite{susskind4}, as well as for a $d=3$ spherically symmetric BTZ-AdS black hole \cite{zhang, btz}.\\
Recently, a novel way for calculating the volume of maximal spacelike hypersurfaces in the bulk in terms of the flux of the volume current was suggested \cite{couch} 

\begin{equation}
\label{eq:11}
V_t=\int_{\Sigma_{t}}v\cdot\epsilon    
\end{equation}
where $\epsilon$ is the spacetime volume element, and $v$ denotes the unit timelike vector field orthogonal to the $t=constant$ hypersurfaces anchored at the boundary. The way we think about $t$ in this context is in terms of asymptotic time, \textit{i.e.} null coordinate which is connected to the Minkowski polar coordinates as $t = r+\tau$ at past infinity, where $\tau$ denotes coordinate time, and $r$ is the radial coordinate \cite{rovelli}. Here, $V(\Sigma_{t})$ is simply the flux of the volume flow through the hypersurface $\Sigma$ at boundary time $t$.\\
The volume current method is valid under the assumption that the whole asymptotically AdS spacetime is foliated (without gaps) by a family of maximal spacelike hypersurfaces (\textit{i.e.} Cauchy slices $\Sigma_{t_{i}}$) induced by the boundary foliation.
As is well known, the CV-duality is intrinsically ill-defined and diverges at asymptotic infinity. This divergence, however, can easily be dealt with by (i) counting only the volume growth within the black hole interior, and (ii) focusing on the volume's time dependence. Following this prescription, one can formulate a pathology-free Second law of complexity and demonstrate that by calculating the volume change between neighboring Cauchy slices behind the horizon in terms of the flow of the volume current,\footnote{Note that in the black hole interior $r<r_{+}$ the volume flow is a future-pointing timelike vector, orthogonal to the maximal foliation.} as we go to later slices, the interior volume can only grow, bounded from above by the complexity of the holographic CFT. 

\subsection{$C^{0}$-instability: What if?}

In light of this, the question we address is: What would a hypothetical validity of Conjecture \hyperlink{conjecture}{1} imply from a quantum complexity-theoretic perspective?\\
As is well known, the inextendibility of the metric to a larger Lorentzian manifold $\tilde{\mathcal{M}}$ beyond $\mathcal{C}\mathcal{H}^{+}$ even as a continuous $C^{0}$ suggests the Cauchy evolution is \textit{spacelike}. Namely, the metric terminates at a singular finite boundary, meaning the region beyond $\mathcal{C}\mathcal{H}^{+}$ does not exist. Essentially, this is the statement that timelike geodesics $\gamma$ cannot reach $\tilde{\mathcal{M}}\setminus\mathcal{M}$.

\begin{figure}
\includegraphics[scale=0.53]{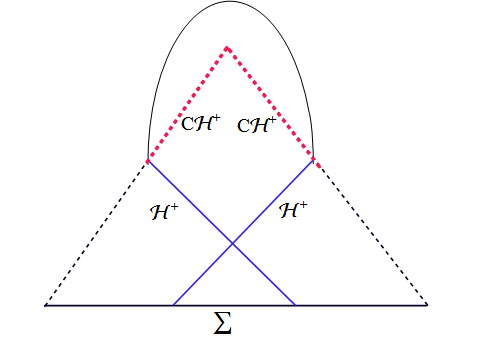}\caption{Maximum Cauchy evolution of asymptotically flat initial data on a spacelike hypersurface $\Sigma$. The region beyond the Cauchy
horizon $\mathcal{C}\mathcal{H}^{+}$ corresponds to the larger Lorentzian manifold. The figure was taken from \cite{dafermos}.}
\label{fig:fig2}
\end{figure}
\begin{figure}
\includegraphics[scale=0.43]{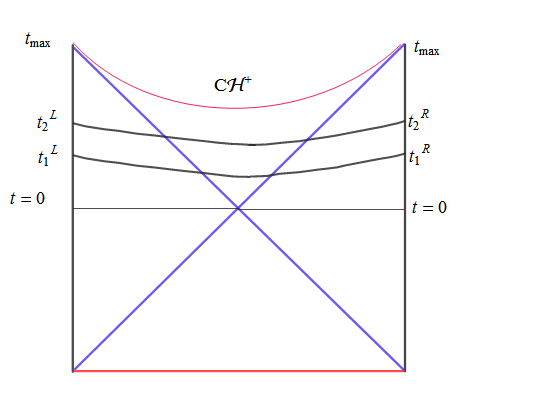}
\caption{A zoomed portion of the upper $t>0$ part (region II from Fig. 1 in \cite{moss}) of a maximally-extended two-sided Kerr-AdS black hole, depicting $\mathcal{C}\mathcal{H}^{+}$ as a migrated singularity. Here, $t_{\text{max}}$ denotes the boundary foliation corresponding to $\mathcal{C}\mathcal{H}^{+}$, where ``max" refers to the latest slice (\textit{i.e.} $\mathcal{C}\mathcal{H}^{+}$), assuming Conjecture \protect\hyperlink{conjecture}{1} is true and the spacetime terminates. Note that $\mathcal{C}\mathcal{H}^{+}$ is depicted as a spacelike boundary for illustrative purposes only; in light of its interpretation as a migrated singularity.}
\label{fig:fig3}
\end{figure}
Loosely speaking, the spacelike finite boundary, corresponding to the maximal Cauchy evolution, could effectively be interpreted as a migrated singularity, where the singularity now lies at some larger $r$, closer to the outer horizon $\mathcal{H}^{+}$.\footnote{In Ref. \cite{susskind5}, examining AMPS' firewall argument, Susskind proposed that black holes which have evaporated less than half of their initial entropy retain their regular horizon structures but have non-standard singularities instead. In particular, it was suggested that the growing von Neumann entropy of the Hawking radiation is what drives the singularity to migrate, and at late times coincide with the event horizon.} This way, by migrating the spacelike finite boundary to coincide or be arbitrarily close to the maximal slice $\Sigma_{\text{max}}$ and $\mathcal{C}\mathcal{H}^{+}$, the Kerr solution obtains the well-defined determinism of the Schwarzschild subfamily.\footnote{For practical purposes we can assume that $\Sigma_{\text{max}}$ and $\mathcal{C}\mathcal{H}^{+}$ coincide.}\\ 
To be more precise, suppose we have a subextremal Kerr-AdS black hole, where we require $M>0$, $a\neq 0$ and $|a|<1/l$

\begin{equation}
\label{eq:kerr-ads}
\begin{split}ds^{2}=-\frac{\Delta_{r}}{\rho^{2}}\left[dt-\frac{a}{\Xi}\sin^{2}\theta d\phi\right]^{2}+\frac{\rho^{2}}{\Delta_{r}}dr^{2}+\frac{\rho^{2}}{\Delta_{\theta}}d\theta^{2}\\
+\frac{\sin^{2}\theta\Delta_{\theta}}{\rho^{2}}\left[adt-\frac{(r^{2}+a^{2})}{\Xi}d\phi\right]^{2}
\end{split}
\end{equation}
where

\begin{equation}
\label{eq:kerr-ads-parameters}
\begin{aligned}\rho^{2}=r^{2}+a^{2}\cos^{2}\theta\\
\Delta_{\theta}=1-l^{2}a^{2}\cos^{2}\theta\\
\Xi=1-l^{2}a^{2}
\end{aligned}
\end{equation}
$l^{2}=-\Lambda/3$ is the AdS curvature radius for negative cosmological constant $\Lambda<0$, and $(r,t,\theta,\phi)\in\mathbb{R}^2\times\mathbb{S}^{2}$ denote the coordinates which cover the black hole interior and the near-horizon region just outside the black hole. Here, the polynomial

\begin{equation}
\label{eq:poly}
\Delta_{r}=(r^{2}+a^{2})(1+l^{2}r^{2})-2Mr\\ 
\end{equation}
vanishes at $\mathcal{H}^{+}$, where $\mathcal{H^{+}}$ is a Killing horizon, and $\frac{M}{\Xi^{2}}$ is the ADM mass. Assuming the weak cosmic censorship, let $(M,a)\in \mathbb{R}^{2}_{>0}$ be non-degenerate for $|a|<1/l$. Then, (\ref{eq:poly}) has 2 real roots for $0<r_{-}<r_{+}$, and the metric has both axial $\partial_{\phi}$ and timelike $\partial_{t}$ Killing fields whose relation to the Killing vector field is the standard

\begin{equation}
\xi=\partial_{t}+\Omega_{\mathcal{H}^{+}}\partial_{\phi}    
\end{equation}
where $\Omega_{\mathcal{H}^{+}}$ is the angular velocity of $\mathcal{H}^{+}$.\\ 
Let us now consider Fig. \ref{fig:fig3}. Given $\mathcal{C}\mathcal{H}^{+}$\footnote{Note that the Cauchy horizon is the surface which separates the space- from timelike character of maximal hypersurfaces $\Sigma_{t}$.} lies at some macroscopic distance from the ringlike curvature singularity $r_{s}$, we get

\begin{equation}
\label{eq:18}
r_{s}<r_{c}<M-\sqrt{M^{2}-a^{2}}    
\end{equation}
where $r_{c}$ denotes the region between $\mathcal{C}\mathcal{H}^{+}$ and the ringlike singularity. Then, assuming the validity of Conjecture \hyperlink{conjecture}{1}, a non-negligible macroscopic region of the black hole interior will, essentially, be cut off.\\
As we show in Section \ref{sec:Section4} below, this interior region cut off is very problematic.

\section{Probing the $C^{0}$-instability}
\label{sec:Section4}

In this Section we study low-temperature black holes with hyperbolic horizons in asymptotically AdS spacetime, dual to CFTs living on hyperboloids. In this setting, we examine the $C^{0}$-form of the SCC from a quantum complexity-theoretic perspective and argue that if Conjecture \hyperlink{conjecture}{1} is true, then the CV-duality would be violated at late times. We demonstrate that the leading contribution to the interior volume comes from a Cauchy slice which lies in the $r_{c}$ region between $r_{s}$ and $\mathcal{C}\mathcal{H}^{+}$, see Eq. (\ref{eq:18}). Given the generic quasi-periodic behavior of quantum complexity, the evolution of the black hole interior will be non-trivially modified. While a macroscopic part of the interior region is cut off, the corresponding complexity  of  the  holographic CFT is expected to continue to evolve until it saturates, hence yielding a deviation from the standard bulk/boundary duality. Therefore, while the complexity of the  boundary CFT continues to grow in time, the dual bulk volume will be prematurely saturated. To preserve the CV-duality, we conjecture that the metric must be $C^{0}$-extendable \emph{at least} for as long as quantum complexity increases. 

The gravitational dual of a CFT on a ($d-1$)-dimensional hyperboloid $H_{d-1}$ is an AdS$_{d+1}$ black hole with hyperbolic horizon geometry. We investigate the $C^{0}$-form of the SCC in the low-temperature limit of these solutions which have been shown to exhibit interesting thermodynamic features. Consequently, we demonstrate that inextendability of the interior metric as a $C^{0}$ beyond $\mathcal{CH}^{+}$ (as suggested by Conjecture \hyperlink{conjecture}{1}) violates the CV-duality as it leads to $\mathcal{O}(1)$ variation in the bulk/boundary correspondence at late times.\\
The hyperbolic AdS$_{d+1}$ black hole has $\mathbb{R}^{2}\times S^{d-1}$ topology and is given by the metric

\begin{equation}
\label{eq:hyperbolicmetric}
ds^{2}=-f(r)dt^{2}+\frac{dr^{2}}{f(r)}+\frac{r^{2}}{l^{2}}d\Sigma^{2}_{d-1}    
\end{equation}
where for the hyperbolic CFT metric

\begin{equation}
f(r)=-1+\frac{r^{2}}{l^2}-\frac{M}{r^{d-2}}    
\end{equation}
and 

\begin{equation}
d\Sigma^{2}_{d-1}=l^{2}dH_{d-1}^{2}    
\end{equation}
where $dH_{d-1}^{2}$ is the unit metric on the $H_{d-1}$ hyperboloid. Here it is convenient to use the convention that lengths are measured in units of the curvature radius (i.e. $l=1$) for simplicity of notation. \\
The black hole's mass $M$ and temperature $T$ are defined, respectively, as 

\begin{equation}
M=r_{0}^{d-2}(r^{2}_{0}-1)    
\end{equation}
and

\begin{equation}
T=\frac{r_{0}}{4\pi}\left(d-\frac{d-2}{r^{2}_{0}}\right)    
\end{equation}
where for $M=0$ we get $T=(2\pi)^{-1}$. In this set of coordinates, $r_{0}$ is the horizon, and in the near-horizon region $f(r)=4\pi T(r-r_{0})$, while in the asymptotic limit, $f(r)\rightarrow r^{2}/l^{2}$.\\
We are interested in low-temperature $T=0$, hyperbolic black holes for which there are solutions that admit regular $r_{-}$ and $r_{+}$ horizons for a family of $-M$ terms.\footnote{Interestingly, $T=0$ hyperbolic black holes have global structure whose interior region resembles that of Reissner-Nordstrom, and has two well-defined horizons, similar to (\ref{eq:kerr-ads}).} We should note that in this respect the negative mass family is special since not all hyperbolic solutions have well-defined black hole configurations, and some have even been shown to break down at $r_{+}$. For instance, the $M=0$ solution only covers part of the manifold (the outside black hole region), and hence is not useful for our considerations.\\
For $T=0$, in the hyperbolic scenario, there are minimum values for $-M$ and $r_{+}$ which are defined, respectively, as

\begin{equation}
\begin{aligned}
&M_{\text{min}}=-\frac{2}{d-2}\left(\frac{d-2}{d}\right)^{d/2}l^{d-2}\\
&r_{\text{min}}=\left(\frac{d-2}{d}\right)^{3/2}l
\end{aligned}
\end{equation}
where for $d=4$ they read

\begin{equation}
\begin{aligned}
&M_{\text{min}}=-\frac{l^{2}}{4}\\
&r_{\text{min}}=\frac{l}{\sqrt{2}}
\end{aligned}
\end{equation}
Motivated by the CV-duality, the volume of a maximal spacelike hypersurface in $T=0$ hyperbolic black hole background can be estimated by dividing the hypersurface $\Sigma_{r_{i}}$ into individual segments, and approximating the local maximum of the radial function $f(r)$ (to be defined below) on each segment. Where in the hyperbolic AdS$_{d+1}$ background (\ref{eq:hyperbolicmetric}), for $r\gg r_{\text{min}}$, the metric asymptotes to an empty AdS$_{d+1}$ solution, while in the near-horizon region, just outside the black hole, the metric is AdS$_{d+1}$ $\times$ $H_{d-1}$.\\
In this setting, each spacelike hypersurface decomposes as\footnote{Although in \cite{barbon} the split was $\Sigma_{t_{i}}=\Sigma_{\text{in}} \cup \Sigma_{\text{out}}$, where $\Sigma_{\text{out}} \sim \Sigma_{\text{R}} \cup \Sigma_{\text{CQM}} \cup \Sigma_{\text{UV}}$, for our analysis we need not consider the details of the exterior components, and will focus only on the interior.}

\begin{equation}
\Sigma_{t_{i}}=\Sigma_{\text{in}} \cup \Sigma_{\text{out}}   
\end{equation}
As is the standard procedure in the literature, the volume of maximal $r=constant$ spacelike hypersurfaces is generally given as

\begin{equation}
r^{d-1}\sqrt{|f(r)|}    
\end{equation}
where $f(r)$ is a ``placeholder," and depends on the region whose contribution to the volume we wish to estimate.\\
For the interior region of (\ref{eq:hyperbolicmetric}), $f(r)$ takes the form \cite{an}

\begin{equation}
\label{eq:maxfunction}
r^{d-1}\sqrt{\left|1-r^{2}+\frac{M_{\text{min}}}{r^{d-2}}\right|}
\end{equation}
and is maximized in the region $r_{c}$, Eq. (\ref{eq:18}). It is important to note that for massive non-extremal black holes the region $r_{c}$ can be arbitrarily big. Therefore, a significant part of the interior volume of low-temperature hyperbolic black holes is contributed by the region $r_{c}$, \textit{i.e.} between $\mathcal{CH}^{+}$ and the ringlike singularity.\\
The overall complexity of this family of solutions has been demonstrated to evolve as \cite{barbon}

\begin{equation}
\label{eq:53}
\Delta\mathcal{C}(t)=\frac{2\alpha}{\sqrt{d}}S\log(lT)^{-1}    
\end{equation}
for small $t$, and at late times as

\begin{equation}
\label{eq:54}
\Delta\mathcal{C}(t)=\frac{2\alpha}{\sqrt{d}}S\log(lT)^{-1}+\frac{4\pi\alpha}{\sqrt{d}}STt   
\end{equation}
Evidently, the growth rate of complexity is consistent with the classical predictions.\\
In this scenario we argue that the local maximum of (\ref{eq:maxfunction}) is very problematic from a complexity-theoretic perspective, where in the context of the SCC, complexity is a measure of the extendability of the metric, \textit{i.e.} the amount of emergent space behind the horizon.\footnote{In a geometric sense, complexity makes a statement about \textit{uniqueness} of the field equations, where considering (\ref{eq:hyperbolicmetric}) and Conjecture \hyperlink{conjecture}{1}, it is used as a probe of the future incompleteness of infalling timelike geodesics.} On the asymptotic AdS$_{d+1}$ boundary the complexity of the holographic CFT, living on a $H_{d-1}$ hyperboloid, as suggested by (\ref{eq:53}, \ref{eq:54}), up to small corrections, grows linearly for time exponential in the entropy until the upper bound is saturated. Meanwhile, in the AdS$_{d+1}$ bulk geometry the dual volume of the interior low-temperature hyperbolic black hole linearly increases, as dictated by the CV correspondence. Assuming the validity of Conjecture \hyperlink{conjecture}{1}, and following the discussion in Section \ref{sec:Section4}, $\mathcal{CH}^{+}$ is a spacelike singular boundary, making timelike geodesics future incomplete.\footnote{The incompleteness is identical to the one found in the interior of a Schwarzschild black hole as an observer asymptotes $r=0$.} That is, the spacetime terminates at $\mathcal{CH}^{+}$ and cannot be extended across, making the region beyond it, effectively, non-existent. In light of (\ref{eq:maxfunction}), however, recall that most of the interior volume of low-temperature hyperbolic black holes is sourced by the region between $\mathcal{CH}^{+}$ and the singularity at $r_{s}$, where (\ref{eq:maxfunction}) maximizes.\\
Hence, as time goes on the bulk volume will saturate much earlier than the complexity of the dual CFT, yielding a discrepancy in the bulk/boundary description which will only accumulate at late times. Therefore, we conjecture that in order to preserve the complexity-theoretic gauge/gravity black hole framework, the interior metric must be extendable to a larger Lorentzian manifold across $\mathcal{CH}^{+}$ (\textit{i.e.} Conjecture \hyperlink{conjecture}{1} must be false) for \textit{at least} a classical recurrence time. Finally, we ought to draw the readers' attention to some conflicting recent results in the literature for different spacetimes. In \cite{kehle, kehle2} the uniform boundedness and the continuity across the Cauchy horizon were demonstrated to remain valid in Reissner-Nordstrom-AdS spacetime for massive scalar waves solving the massive Klein-Gordon equation, assuming a sharp logarithmic decay rate. Despite these promising results invalidating the $C^{0}$ SCC, however, the question of boundedness remained open for other general spacetimes. On the other hand, a rigorous proof was presented recently \cite{kehle3}, suggesting that if the black hole's mass and angular momentum satisfy some non-Diophantine conditions, then linear scalar perturbations (solving the Klein-Gordon equation) blow up at the Cauchy horizon. Thus, pointing towards a positive resolution of the $C^{0}$ SCC for a Kerr-AdS black hole, given the parameters admit Baire genericity. Interestingly, this instability was shown not to result from mere superradiance or blueshift effects but instead from a novel resonance phenomenon (\textit{i.e.} relation between Diophantine approximation and black hole interiors) found in axisymmetric solutions.\\  

\section{Conclusions}

In this work we presented a novel quantum treatment of an old problem concerning metric extendability in black hole interiors with inner horizons. This is reminiscent of the early quantum approach to black holes which led to completely redefining the subject in the 70s. We employed a quantum information framework which has been recently shown to play a key role in gravity and spacetime emergence, and to our knowledge, has never been considered in this context.\\
In particular, we revisited, from a quantum complexity-theoretic perspective, the $C^{0}$ formulation of the SCC in the low-temperature limit of a hyperbolic AdS$_{d+1}$ black hole dual to a CFT living on a $(d-1)$-dimensional hyperboloid $H_{d-1}$. Our analysis showed that inextendability of the metric across $\mathcal{CH}^{+}$ (validity of Conjecture \hyperlink{conjecture}{1}) would lead to late-time violation of the CV-duality, as most of the interior black hole volume is sourced from the region beyond $\mathcal{CH}^{+}$. To retain the CV-duality and save the bulk/boundary correspondence, we conjectured imposing a lower bound on metric extendability of order the classical recurrence time.\\

\section*{Acknowledgments}

AY is grateful to the University of Oxford, where part of the work was conducted, for the hospitality. VV thanks the Oxford Martin School, the John Templeton Foundation, and the EPSRC (UK).

\bibliographystyle{plain}

\end{document}